\documentclass[aps,pra,preprint,floatfix]{revtex4-1}% Physical Review A

\usepackage{graphicx}
\usepackage{amsmath}
\usepackage{amssymb}
\usepackage{bm}
\usepackage{color}

\def\beq{\begin{equation}}
\def\eeq{\end{equation}}

\def\om{\omega}
\def\a{\alpha}
\def\b{\beta}

\def\eps{\epsilon}

\def\l{\lambda}
\def\s{\sigma}
\def\D{\Delta}

\def\ad{a^\dagger}
\def\rd{{\rm{d}}}

\def\p{\phi}

\def\ra{\rightarrow}

\def\Cc{\mathbb{C}}
\def\Zz{\mathbb{Z}}
\def\Rr{\mathbb{R}}

\def\dd{\frac{\rd}{\rd z}}

\def\ua{\!\!\uparrow}
\def\dd{\frac{\rd}{\rd z}}
\def\da{\!\!\downarrow}
\def\dd{\frac{\rd}{\rd z}}

\begin{document}

\title[Comment on `Parameter-dependent unitary transformation approach for quantum Rabi model']
{Comment on `Parameter-dependent unitary transformation approach for quantum Rabi model'}

\author{Daniel Braak$^1$, Murray T. Batchelor$^2$ and Qing-Hu Chen$^3$}

\address{$^1$Department of Physics, Augsburg University, 86159 Augsburg, Germany}

\address{$^2$Mathematical Sciences Institute, Australian National University, Canberra ACT 2600, Australia}

\address{$^3$Department of Physics, Zhejiang University, Hangzhou 310027, China}

\begin{abstract}
\noindent
We discuss the elementary errors in D.G.~Zhang's \cite{zhang} claimed exact solution of the quantum Rabi model.
The erroneous solution is seen to be nothing more than the combined solution of the simpler Jaynes-Cummings and
anti-Jaynes-Cummings models obtained by neglecting terms in the model Hamiltonian.
\end{abstract}

\maketitle

%\noindent
%Comment on `Parameter-dependent unitary transformation approach for quantum Rabi model', D.G.~Zhang, \textit{New J. Phys.} \textbf{23}, 093014 (2021)\par
%\vspace{5mm}
%\noindent

Degang Zhang \cite{zhang} studies the asymmetric quantum Rabi model with Hamiltonian
\beq
H_R=\om\ad a +g(a+\ad)\s_x +\l\s_z +\eps\s_x.
\label{QRM}
\eeq
This model has been solved analytically in \cite{braak_11,chen_12,zhong_13,MPS_14}, see also the review \cite{xie_17}.
Zhang claims that this solution is incorrect and presents an alternative solution.
To simplify the argument, we shall concentrate in the following on the case $\eps=0$, where the quantum Rabi model (QRM) possesses a discrete $\Zz_2$-symmetry.
The spectral graph consists therefore of two intersecting ladders (``subspectra'') with defined parity $\pm 1$, the eigenvalues of the symmetry operator.
Zhang does not obtain this spectral structure but finds instead four subspectra.
The energies of Zhang's eigenstates are given in closed form as
\begin{align}
  E_{n,\pm}^I=&\om\left(n+\frac{1}{2}\right)\pm\sqrt{\left(\frac{\om}{2}+\l\right)+(n+1)g^2},\qquad\textrm{subspectrum I},
  \label{sub1}\\
  E_{n,\pm}^{II}=&\om\left(n+\frac{1}{2}\right)\pm\sqrt{\left(\frac{\om}{2}-\l\right)+(n+1)g^2},\qquad\textrm{subspectrum II},
  \label{sub2}
\end{align}
for $n=0,1,2\ldots$.
These are the main results (Z26) and (Z13) in Zhang's paper.
However, a simple textbook calculation reveals that these spectra are precisely those of the well known Jaynes-Cummings model (subspectrum II)
\beq
H_{JC}=\om\ad a +g(a\s^++\ad\s^-) +\l\s_z,
\eeq
and the anti-Jaynes-Cummings model (subspectrum I)
\beq
H_{AJC}=\om\ad a +g(a\s^-+\ad\s^+) +\l\s_z.
\eeq

\begin{figure}[tbph]
\centering
\includegraphics[width=0.49\linewidth]{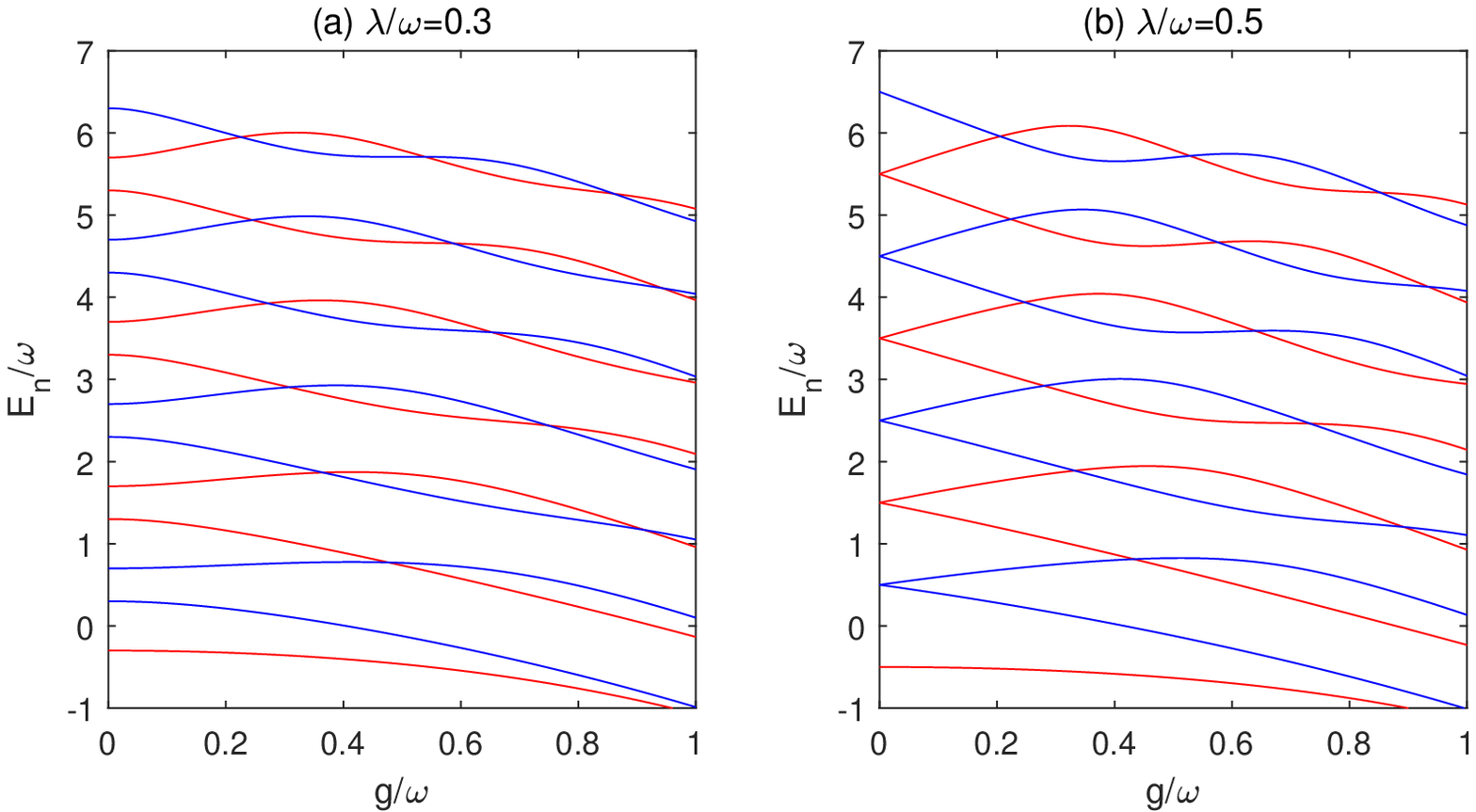}
\includegraphics[width=0.49\linewidth]{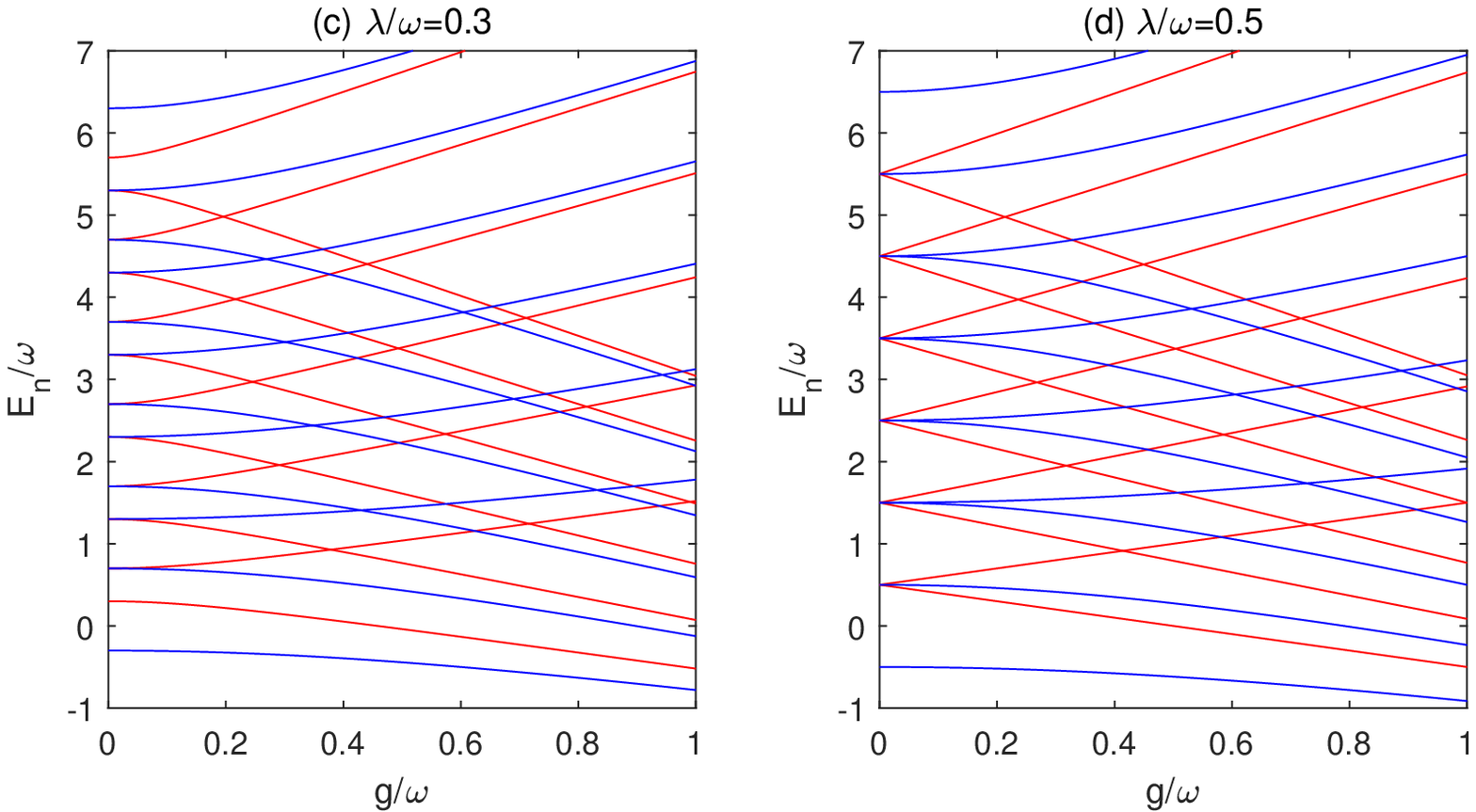}
\caption{
The plots in figures (a) and (b) show the lowest energy levels of the QRM as a function of the coupling $g/\omega$.
The parameter values $\lambda/\omega=0.3$ and $\lambda/\omega=0.5$ are chosen to be the same as Zhang's Figure 1.
The red and blue lines correspond to the two parities $+1$ and $-1$, respectively.
For comparison, Zhang's spectrum, comprising the energy levels of the JC model Eq.~(\ref{sub2}) and the AJC model Eq.~(\ref{sub1}) 
are shown in figure (c) and (d) for the same parameter values. Here the JC results are shown in red and the AJC results in blue.
%For each case the solid (dashed) lines correspond to the + (-) sign in Eqs.~(\ref{sub1}) and (\ref{sub2}).
The latter results are erroneously claimed to be the energy spectrum of the QRM. }
\label{comparison}
\end{figure}

In Figure~\ref{comparison} we compare Zhang's eigenspectrum with the exact numerical results for the QRM.
As to be expected, the difference is clear.

For both the JC and AJC models, the Hilbert space $L^2(\Rr)\otimes\Cc^2$ separates into infinitely many two-dimensional invariant subspaces,
because the symmetry is enhanced from $\Zz_2$ to $U(1)$.
In the case of the QRM, the Hilbert space separates into  only two invariant subspaces, each infinite-dimensional.
It is therefore obvious that the solution presented by Zhang must be incorrect, as it merely reproduces the combined spectrum of the JC and the AJC models.
Indeed, the error is  rather elementary.
In the language of spectral decomposition, the spectrum of \eqref{QRM} is a pure point spectrum,
because the operator $\l\s_z+g(a+\ad)\s_x$ is norm bounded relatively to $\om\ad a$.
Therefore, all eigenstates must be normalizable.
Zhang makes the correct general ansatz for the eigenfunction $\bm{\psi}$ in $L^2(\Rr)\otimes\Cc^2$ as an infinite series in the oscillator eigenstates
$\{|m,\,\da\rangle, |m,\, \ua\rangle\}$ for $m=0,1,2,\ldots$, Eq.~(Z2) in the paper (according to Zhang, we may set $\D_{ns}=0$).
This state will be normalizable if
\beq
|\bm{\psi}|^2=\sum_{m=0}^\infty|\a_m|^2+|\b_m|^2 <\infty.
\label{norm}
\eeq
Plugging the ansatz into the Schr\"odinger equation $H_R\bm{\psi}=E\bm{\psi}$, Zhang obtains the three-term  recurrence relations Eqs.~(Z3) and (Z4),
containing the energy $E$ as parameter.
These recurrence relations determine the coefficients $\a_m,\b_m$ for $m\ge 1$, starting from $\a_0,\b_0$ in terms of $E$.
In some cases, including the JC and the AJC models, there is a solution for $\{\a_m,\b_m\}$ where only a finite number of coefficients are non-zero.
These states are automatically normalizable.
The situation is different for the QRM, where all formal solutions of the Schr\"odinger equation defined by the recurrence relations are \emph{infinite} series in terms of oscillator eigenstates.
The state $\bm{\psi}$ is only normalizable for a discrete set of values of the parameter $E$, the eigenenergies of $H_R$.
These energies are therefore determined by the condition \eqref{norm} above.
It is necessary to recall these elementary facts here because Zhang is apparently unaware of them.
Instead of implementing \eqref{norm}, which is never mentioned in the paper, two arbitrary conditions are imposed on the parameters $\a_{n+1}, \b_n$, Eqs.~(Z5) and (Z6),
for subspectrum I, respectively Eqs.~(Z18) and (Z19) on $\a_n,\b_{n+1}$ for subspectrum II.
From these conditions, the parameters $E_{n,\pm}^I$ respectively $E_{n,\pm}^{II}$ are obtained.
Neither Eqs.~(Z5) and (Z6) nor (Z18) and (Z19) have anything to do with the spectral condition \eqref{norm} --
they do not guarantee the vanishing of all coefficients $\a_m,\b_m$ for $m>n+1$ or any other property of the set $\{\a_m,\b_m\}$ entailing the finiteness of $|\bm{\psi}|^2$.
Zhang acknowledges the fact that the series for $\bm{\psi}$ does not terminate in his Eq.~(Z16) but says nothing about the convergence of the series in \eqref{norm}.

The states $\bm{\psi}(E)$ with energy parameter of the form (\ref{sub1}) and (\ref{sub2}) are thus not normalizable and those values for $E$ do not form the QRM spectrum.
Apart from the obvious senselessness of conditions (Z5), (Z6), (Z18) and (Z19) there is a very simple way to prove this analytically by employing the quasi-exact (Juddian) solutions of the QRM \cite{judd_79,kus_85,kus_86}.
These exact eigenstates have the simplest form  in the Bargmann space of analytic functions.
In this representation the Schr\"odinger equation is equivalent to the set of coupled first order differential equations
\begin{align}
  (z+g)\dd\p_1 +gz\p_1 +\l\p_2=&E\p_1,
  \label{r1}\\
  (z-g)\dd\p_2 -gz\p_2 +\l\p_1=&E\p_2,
  \label{r2}
\end{align}
for $\om=1$ (see Eqs.~(ZA3) and (ZA4) in Zhang's paper). Now it is easy to check that $\bm{\psi}=(\p_1,\p_2)^T$ with
\beq
\p_1(z)=e^{-gz}\left(\frac{2g}{\l}z+\l+\frac{2g^2}{\l}\right), \quad
\p_2(z)=e^{-gz},
\label{judd}
\eeq
solves (\ref{r1}) and (\ref{r2}) with energy $E=1-g^2$ if $g$ and $\l$ satisfy the condition $\l^2+4g^2=1$.
The question is now whether the exact eigenvalue $1-g^2$ appears among the sets (\ref{sub1}) and (\ref{sub2}) for some integer number $n$ if $\l=\sqrt{1-4g^2}$ and $\om=1$.
This is not the case: setting, e.g., $\l=1/2$, $g=\sqrt{3}/4$ one finds $n\simeq0.714$ (subspectrum I) and $n\simeq0.0986$ (subspectrum II).

The functions $\p_1(z),\p_2(z)$ in \eqref{judd} are analytic in $z$ and normalizable with respect to the Bargmann norm (Eq.~(Z29)). They correspond thus to an eigenstate of the QRM,
although they diverge on the real axis for $z\ra-\infty$.
They have a form similar to the functions in Eqs.~(ZA5) and (ZA6) for which Zhang claims are not normalizable due to this divergence.
Hand in hand with the earlier errors, Appendix A of Zhang's paper demonstrates a lack of working knowledge of basic complex analysis and the theory of ordinary differential equations.
The argument used in \cite{braak_11} is explained in considerable detail in \cite{braak_13_genG}.
Moreover, it has been rigorously demonstrated in  \cite{braak_13_contfrac} that the numerical evaluation of the QRM spectrum via exact diagonalization in a truncated Hilbert space, shown in Figs.~\ref{comparison}(a),(b), converges to the exact solution.
It should be further noted that Zhang's claim not only contradicts the exact results in \cite{braak_11,chen_12,zhong_13,MPS_14}
but all known numerical and analytical approaches to the quantum Rabi model which have been established over many years, see, e.g.,  \cite{feranchuk_96,irish_07,gan_10,pan_17,kockum_19,forn-diaz_19,le_boite_20}.

\bibliography{NJPcomment.bib}   % name your BibTeX data base

\end{document}